\documentstyle[twoside,psfig]{article}
\input{ijmpb-macros}

\begin{document}

\runninghead{Itinerant and localized states
in strongly correlated systems \ldots
}
{Itinerant
and localized states in strongly correlated systems \ldots
}

\normalsize\textlineskip
\thispagestyle{empty}
\setcounter{page}{1}

\copyrightheading{}			

\vspace*{0.88truein}

\fpage{1}
\centerline{\bf ITINERANT AND LOCALIZED STATES}
\vspace*{0.035truein}
\centerline{\bf IN STRONGLY CORRELATED SYSTEMS BY A MODIFIED}
\vspace*{0.035truein}
\centerline{\bf MEAN-FIELD SLAVE-BOSON APPROACH}
\vspace*{0.37truein}

\centerline{\footnotesize EMMANUELE CAPPELLUTI}
\vspace*{0.015truein}
\centerline{\footnotesize\it Dipartimento di Fisica, 
Universit\`a ``La Sapienza'', P.le Aldo Moro 2}
\baselineskip=10pt
\centerline{\footnotesize\it Roma, 00185, Italy} 
\baselineskip=10pt
\centerline{\footnotesize\it and INFM, Unit\`a Roma1}
\vspace*{0.225truein}

\abstracts{The standard mean field slave-boson solution for the infinite-$U$
Hubbard model is revised. A slightly modified version is proposed
which includes properly the incoherent contribution of the localized states.
In contrast to the standard mean field result, this new proposed
solution defines a unique spectral function to be used in the calculation of
local and not local quantities, and satisfies the correct thermodynamic
relations. The same approach is applied also to the mean field approximation
in terms of Hubbard operators. As a byproduct of this analysis,
Luttinger's theorem is shown to be fulfilled in a natural way.
}{}{}

\textlineskip			
\vspace*{12pt}			

\textheight=7.8truein
\setcounter{footnote}{0}
\renewcommand{\thefootnote}{\alph{footnote}}

\section{Introduction}

A paradigmatic model to study systems of strongly correlated electrons is
the Hubbard model.
It contains the minimum of features necessary 
to describe bandlike itinerant or localized electrons depending
on microscopic parameters. The Hamiltonian has the simple form:
\begin{equation}
H = - \sum_{\sigma} \sum_{i j} t_{i j} c^\dagger_{i \sigma} c_{j \sigma}
+ U \sum_i c^\dagger_{i \uparrow} c_{i \uparrow}
c^\dagger_{i \downarrow} c_{j \downarrow},
\label{hub}
\end{equation}
where the first term is a one-band tight-binding
Hamiltonian. Its first kinetic term describes the destruction of an electron 
on site $j$ and the creation of an electron
with the same spin on site $i$. The second term in equation (\ref{hub})
is the on-site repulsion between electrons, which
is expected to be relevant in real materials for $d$- and $f$- orbitals.
While the limiting case $U \ll t$ can easily be dealt with by means
of perturbation theory, leading to a renormalized band of quasiparticles,
the intermediate and the strong-coupling limits $U \simeq t$ are much more 
interesting.
From a qualitative point of view, in these regimes
the system is described by a narrow coherent band of
itinerant quasiparticles on the top of a large background of incoherent
localized states.
The metal-insulator transition is thus characterized by the disappearing
of the coherent band.
This picture is confirmed by dynamical mean field analyses\cite{georges}.
In the strong coupling case ($U \gg t$) the bandwidth and
the spectral weight of the
itinerant quasiparticles scales only with the number of holes $\delta$,
so that an insulating state is achieved at half-filling $\delta=0$.
In such a situation the total weight of the spectral function 
is in its incoherent part.

On the analytical ground, all the possible informations about
the single-particle properties of the system are obtained by the
knowledge of the one-electron Green's function $G$.
Without losing any generality, it can be splitted in a coherent and
an incoherent contribution\cite{abrikosov}:
\begin{equation}
G_\sigma({\bf k},\omega) =
\frac{Z}{\omega-\epsilon({\bf k})+\mu+i0^+\: i{\rm sgn} [\epsilon({\bf k})-\mu]}
+G^{{\rm inc}}_\sigma({\bf k},\omega),
\end{equation}
where $Z$ is the spectral weight of itinerant states with
effective dispersion $\epsilon({\bf k})$
and $G^{{\rm inc}}$
contains all the physics not described by the first term.
For the above arguments, we expect $Z \propto \delta$.

A meaningful quantity that can be calculated by the Green's function
is the occupation number
\begin{equation}
n_\sigma({\bf k})= \int 
\frac{d\omega}{2\pi i} G_\sigma({\bf k},\omega)
{\rm e}^{-i\omega 0^-}.
\end{equation}
In a similar way as the Green's function, $n_\sigma({\bf k})$ can be
distinguish in a coherent part, with a sharp jump of height $Z$
at the Fermi surface, plus the incoherent background spread over
the whole Brillouin zone. 

One of the most popular tools to deal with strongly correlated
systems is the slave-boson technique\cite{coleman}.
Its feasibility makes it
suitable to be applied to different models, as for instance
the Kondo problem or the Anderson Hamiltonian\cite{bickers}.
In the finite $U$ Hubbard model, it is applied by introducing four auxiliary
bosonic fields\cite{kotliar1}, while in the $t$-$J$ Hamiltonian, obtained
as strong-coupling limit of the Hubbard one, only
one boson is needed.
For the purpose of this work we restrict our study
to the simplest case of the so-called ``$t$-model'', equivalent
to the infinite-$U$ Hubbard model or to the $t$-$J$ model with
zero exchange coupling constant.
In this case the strong correlation arises from the constraint
of no double occupancy on each site.

As above discussed, in the $U \rightarrow \infty$ limit
the metal-insulator transition, at zero temperature, is driven
by only one relevant parameter, the hole number $\delta$,
related to the total number of electrons $n$ by $\delta=1-n$.
In the mean-field approximation,
which represents its simplest formulation, the slave-boson
technique maps the strong correlated Hamiltonian into
an effective model of non interacting fermions with
renormalized bandwidth\cite{kotliar2}. 
Thus, the resulting Hamiltonian is thought
to describe the physics of the itinerant coherent states, whereas
the contribution of the localized states is totally disregarded
at this level, and it can be restored only by higher order
approximations. In agreement with this picture, the spectral weight
associated with the coherent states can be shown to be equal
to the doping $\delta$ and it can be therefore identified with $Z$.
If one considers the occupation number of the electrons,
the mean-field solution of the slave-boson approach accounts
only for the sharp jump at the Fermi energy while the
incoherent background is missing (figure 1).
Incoherent states are described in the slave-boson picture only at
higher order than the mean-field solution.
An analysis of the finite-$U$ Hubbard model
shows that the dynamics of the auxiliary bosons
reconstructs the lower and upper Hubbard bands with splitting 
$\sim U$\cite{raimondi}. However in the infinite-$U$ Hubbard model,
where single occupancy constraint is more compelling and no energy scale
but the kinetic one exists, it is not clear where the incoherent background
should be located and how the total sum rule fulfilled.
In the $t$-$J$ model, where the inclusion of the exchange term identifies
a characteristic magnetic energy, spectral weight is expected
to arise on a scale $J$ associated with spin excitations.
However, the violation or less of the total sum rule
is basically connected with the single occupancy constraint, independently
of the exchange term which would just redistribute the total spectral weight.
In this perspective, the simple
$t$-model is as respresentative as the $t$-$J$ in order to test
the conservation of the total spectral weight.

\begin{figure}[t]
\centerline{\psfig{figure=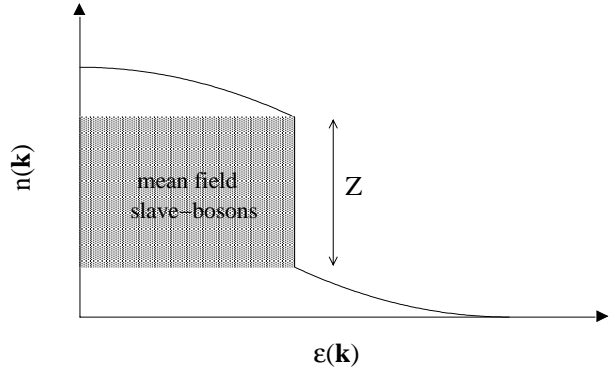,width=8cm}}
\caption{Sketch of the electron occupation number. The shaded area
represents the spectral weight of the itinerant quasi-particles
as described by the mean-field slave-boson solution.}
\end{figure}

The aim of this paper is to revise the mean-field solution
in order to take into account in a proper way the incoherent background.
It is shown that, even in mean-field approximation,
a rigorous implementation of the no double occupancy constraint
leads to an analytic expression of the Green's function
that describes on the same foot itinerant and localized states.
According to its mean-field nature
the proposed approach shares with the standard mean-field
solution the common shortcomings due to neglecting the boson dynamics.
In particular incoherent states in lower and upper bands can not
be described. Nevertheless it provides 
an improved version of the mean-field slave-boson solution
free of some inconsistencies of the standard one and preserving
the correct spectral weight sum rule.
The same procedure is applied to the
``mean-field'' solution of the $t$-model in terms of Hubbard $X$-operators, 
which differs deeply
from the one obtained by the slave-boson technique.
It is shown that the inclusion of the incoherent part permits
to overcome some intrinsic inconsistencies of the $X$-operator solution,
as for instance about the validity of Luttinger's theorem.

\section{Mean field approximation in the slave-boson approach}

Let us start with a brief summary of the well-known mean field
solution within the slave-boson formalism\cite{coleman}. 
This reviewing is finalized to present all the main analytic steps
of such a derivation permitting to point out the critical passages
that will be afterwards modified in order to obtain a consistent 
picture.

Let us consider the Hamiltonian of the $t$-model:
\begin{equation}
H=\sum_{i \neq j,\sigma} t_{ij} 
\tilde{c}^\dagger_{i\sigma}\tilde{c}_{j\sigma}
-\mu\sum_{i,\sigma} \tilde{c}^\dagger_{i\sigma}\tilde{c}_{i\sigma},
\label{hc}
\end{equation}
where $\tilde{c}$, $\tilde{c}^\dagger$ 
are electronic fields operating on the reduced Hilbert space
with no double occupied sites.

In the slave-boson formalism
the creation (destruction) operator of the real electron
is decomposed in two operators by the usual substitution:
\begin{eqnarray}
\tilde{c}^\dagger_{i\sigma} &\longrightarrow& b_i f^\dagger_{i\sigma},
\label{cbf1}\\
\tilde{c}_{i\sigma} &\longrightarrow& b^\dagger_i f_{i\sigma},
\label{cbf2}
\end{eqnarray}
where the $b$ and $f$ operators fulfil respectively bosonic
and fermionic algebras.
The constraint of no double occupancy is implemented
on each site by the condition
\begin{equation}
b^\dagger_i b_i +f^\dagger_{i\uparrow}f_{i\uparrow}
+f^\dagger_{i\downarrow}f_{i\downarrow}=1.
\label{constraint}
\end{equation}

By using the relations (\ref{cbf1})-(\ref{cbf2}), and introducing
the Lagrange multiplier $\lambda_i$ on each site to enforce
the constraint, the slave-boson Hamiltonian reads:
\begin{equation}
H=\sum_{i \neq j,\sigma} t_{ij} 
b_i b^\dagger_j f^\dagger_{i\sigma}f_{j\sigma}
-\mu\sum_{i,\sigma} f^\dagger_{i\sigma}f_{i\sigma}
+\sum_i\lambda_i\left[b^\dagger_i b_i +f^\dagger_{i\uparrow}f_{i\uparrow}
+f^\dagger_{i\downarrow}f_{i\downarrow}-1\right].
\label{hbf}
\end{equation}
In the derivation of equation (\ref{hbf}) the relation
\begin{equation}
\tilde{c}^\dagger_{i\sigma}\tilde{c}_{i\sigma}=
b_i b^\dagger_i f^\dagger_{i\sigma}f_{i\sigma}
= f^\dagger_{i\sigma}f_{i\sigma}
\label{bfi}
\end{equation}
has been used.

In the mean-field approximation 
the local fields $b_i(t)$, $b^\dagger_i(t)$, $\lambda_i(t)$ are replaced
by their mean-field global values:
$b_i(t) \rightarrow \langle b_i \rangle \equiv b$, 
$b^\dagger_i(t) \rightarrow  \langle b^\dagger_i \rangle \equiv b^*$,
$\lambda_i(t) \rightarrow \langle \lambda_i \rangle \equiv \lambda$.
Note that in such an approximation both the space and time
dependences of $b$ and $\lambda$ are dropped.
The resulting Hamiltonian describes an effective
model of renormalized non interacting fermions:
\begin{equation}
H^{MF}=\sum_{i \neq j,\sigma} b^2 t_{ij} f^\dagger_{i\sigma}f_{j\sigma}
-(\mu-\lambda)\sum_{i,\sigma} f^\dagger_{i\sigma}f_{i\sigma}
+\lambda\sum_i\left[b^2 -1\right],
\label{hbfmf}
\end{equation}
or, in Fourier space,
\begin{equation}
H^{MF}=\sum_{{\bf k},\sigma} 
\left[ b^2 t({\bf k}) -\mu +\lambda\right]
f^\dagger_{{\bf k}\sigma}f_{{\bf k}\sigma}
+\lambda \left[ b^2 -1 \right].
\label{hbfmfk}
\end{equation}

The Green's function of the $f$-fermions takes the simple form
\begin{equation}
G^f_\sigma({\bf k};\omega)=\frac{1}{\omega-b^2t({\bf k})+\mu-\lambda
+\: i{\rm sgn} [b^2t({\bf k})-\mu+\lambda]},
\label{gf}
\end{equation}
that represents the Green's function of purely itinerant fermions
with bandwidth renormalized by a factor $b^2$
and chemical potential shifted by $-\lambda$.
The corresponding zero temperature occupation number of the $f$-particles,
\begin{equation}
n^f_\sigma({\bf k})=\int \frac{d\omega}{2\pi i} G^f_\sigma({\bf k},\omega)
{\rm e}^{-i\omega 0^-},
\end{equation}
has thus a jump from 1 to 0 at the chemical potential.
For simplicity in the following the zero temperature case will be 
always considered. Discussion and results can be however
straightforwardly generalized
at finite temperature.

The internal energy can be
expressed as function of $n^f_\sigma({\bf k})$:
\begin{equation}
E=\sum_{{\bf k},\sigma} 
\left[ b^2 t({\bf k}) -\mu +\lambda\right]
n^f_\sigma({\bf k})
+\lambda \left[ b^2 -1 \right].
\label{freenfk}
\end{equation}
The physical solution for the mean-field parameters $b^2$ and $\lambda$
is obtained by minimizing equation (\ref{freenfk}).
Their analytical expressions are:
\begin{equation}
\lambda = -\sum_{{\bf k},\sigma} t({\bf k})
n^f_\sigma({\bf k}),
\end{equation}
and
\begin{equation}
b^2=1-\sum_{{\bf k},\sigma} n^f_\sigma({\bf k}).
\end{equation}
By using these relations, equation (\ref{freenfk}) is therefore simplified:
\begin{equation}
E=\sum_{{\bf k},\sigma} 
b^2 t({\bf k}) n^f_\sigma({\bf k})
-\mu \sum_{{\bf k},\sigma}
n^f_\sigma({\bf k}).
\label{freenfk2}
\end{equation}

The total number of electrons $n$ is easily obtained by
the thermodynamic relation $n = -\partial E / \partial \mu$:
\begin{equation}
n=n^f=\sum_{{\bf k},\sigma} n^f_\sigma({\bf k}),
\label{nnf}
\end{equation}
which identifies the number of ``real'' electrons with the number 
of the $f$-fermions, that is to say with the number of single occupied
states. 

Equation (\ref{nnf}) is indeed consistent with
relation (\ref{bfi}). 
However, some intrinsic inconsistencies appear just by lo\-ok\-ing
at the spectral function.
If one considers the Green's function of real electrons
\begin{equation}
G_\sigma(i,j;t)=
-i\langle T_t c_{i\sigma}(t)c^\dagger_{j\sigma}\rangle,
\label{gc}
\end{equation}
the relation between $G_\sigma$ and $G^f_\sigma$,
within the mean field approximation,
is simply:
\begin{equation}
G_\sigma(i,j;t)= b^2 G^f_\sigma(i,j;t).
\label{gcgf}
\end{equation}
The same relation thus holds for the spectral functions
$A({\bf k},\omega)= b^2 A^f({\bf k},\omega)$, 
and for the corresponding electronic occupation number
\begin{equation}
n({\bf k})=b^2 n^f({\bf k}).
\label{nknfk}
\end{equation}

This result is in partial agreement with
the physical situation depicted in figure 1,
where the contribution of the coherent states to the total number of electrons
scales with $Z=\delta=1-n$, but it is in open contrast with
Equation (\ref{nnf}). 
In order to reconcile the two descriptions one should assume
two relations linking $n({\bf k})$ with $n^f({\bf k})$, respectively
$n({\bf k})=n^f({\bf k})$ and $n({\bf k})=b^2n^f({\bf k})$.
The first one should be employed to calculate only {\em local}
quantities as the total number of electrons $n$, as it has implicitly done
in the expression of the internal energy in equation (\ref{freenfk2}).
The second one should be used for {\em non local} quantities, as
shown by the kinetic energy in the same equation.
It is not clear in this procedure which of them is the 
{\em physical} occupation number of the real electrons.

The inconsistency is clearly related to the neglecting of the incoherent
background of localized states.
The spectral function $A({\bf k},\omega)$ as well as the
occupation number $n({\bf k})$ of the real electrons should contain
both the coherent and incoherent contributions.
In the following it is shown how the previous approach can be
modified to take into account in a simply way both the coherent
and incoherent parts, leading to an unambiguous determination
of the spectral function and of the related quantities $n({\bf k})$ and $n$.

\section{Modified mean-field approach}

Actually, the procedure here proposed is quite simple.
The basilar consideration arises from the
realizing that the above discrepancies
stem from the different implementations of the no double occupancy constraint.
Indeed, in the evaluation of the local quantity 
$\tilde{c}^\dagger_{i\sigma}\tilde{c}_{i\sigma}$ the condition
(\ref{bfi}) has been used to identity
$\langle\tilde{c}^\dagger_{i\sigma}\tilde{c}_{i\sigma}\rangle$
with $\langle f^\dagger_{i\sigma}f_{i\sigma}\rangle$.
A similar equality however does not hold for the 
non local Green's function
of the real electrons defined in equation (\ref{gc}).
In this situation equation (\ref{bfi}) can not be applied whereas
the relation (\ref{gcgf}) is rather fulfilled.
An appropriate way to deal with the generic Green's function is to
split it in the local and non local part:
\begin{equation}
G_\sigma(i,j;t)=-i\langle  T_t
\tilde{c}_{i\sigma}(t)\tilde{c}^\dagger_{j\sigma}\rangle
\left[1-\delta_{ij}\right]
-i\langle T_t\tilde{c}_{i\sigma}(t)
\tilde{c}^\dagger_{i\sigma}\rangle\delta_{ij}.
\label{gsplit}
\end{equation}

Let us now introduce the slave-boson formalism and perform
the mean field approximation in two steps: firstly
the fields $b_i(t)$, $b^\dagger_i(t)$ are replaced by their
mean field values with respect to the time 
$b_i(t) \rightarrow b_i$, $b^\dagger_i(t) \rightarrow b^\dagger_i$.
Note that they still preserve a full site dependence.
As a consequence of this first approximation the relation
(\ref{bfi}) can now be formulated in the dynamic version:
\begin{equation}
\tilde{c}^\dagger_{i\sigma}(t)\tilde{c}_{i\sigma}=
b_i b^\dagger_i f^\dagger_{i\sigma}(t)f_{i\sigma}
= f^\dagger_{i\sigma}(t)f_{i\sigma}.
\label{bfit}
\end{equation}

We can now safely apply in equation (\ref{gsplit})
the spatial mean field approximation in the first non local part, while
in the second local term the relation (\ref{bfit}) can correctly
be employed. The resulting Green's function becomes:
\begin{equation}
G_\sigma(i,j;t)=-ib^2\langle T_t
f_{i\sigma}(t)f^\dagger_{j\sigma}\rangle
\left[1-\delta_{ij}\right]
-i\langle T_t f_{i\sigma}(t)f^\dagger_{i\sigma}\rangle\delta_{ij},
\end{equation}
or, in a more compact way,
\begin{equation}
G_\sigma(i,j;t)=-ib^2\langle T_t
f_{i\sigma}(t)f^\dagger_{j\sigma}\rangle
-i\left[1-b^2\right]\langle T_t
f_{i\sigma}(t)f^\dagger_{i\sigma}\rangle\delta_{ij}.
\end{equation}

In Fourier space, the analytic expression of the electronic
Green's function takes the form:
\begin{equation}
G_\sigma({\bf k};\omega)= Z G^f_\sigma({\bf k};\omega)
+\left(1-Z\right) \sum_{{\bf k}} G_\sigma({\bf k};\omega),
\label{ggl}
\end{equation}
where $Z$ is the coherent spectral weight $Z=b^2$.
A straight\-for\-ward consequence of equation (\ref{ggl}) is:
\begin{equation}
\sum_{{\bf k}} G_\sigma({\bf k};\omega)=
\sum_{{\bf k}} G^f_\sigma({\bf k};\omega),
\label{glgfl}
\end{equation}
so that the electronic propagator $G$
can be fully expressed as function of $G^f$:
\begin{equation}
G_\sigma({\bf k};\omega)= Z G^f_\sigma({\bf k};\omega)
+\left(1-Z\right) \sum_{{\bf k}} G^f_\sigma({\bf k};\omega).
\label{ggfl}
\end{equation}

From equation (\ref{ggl}), the occupation number of the real electrons
is also immediately evaluated:
\begin{equation}
n_\sigma({\bf k}) = Z n^f_\sigma({\bf k})+\left(1-Z\right) n_\sigma,
\label{nkr}
\end{equation}
where $n_\sigma$ is the total number of electrons per spin.

The analytic expressions of the Green's function for the $f$-fermions,
and con\-se\-quent\-ly of $n^f_\sigma({\bf k})$,
in the present approach are just the same as in the standard
mean field slave-boson result given by equation (\ref{gf}).
This can be easily checked by considering that the mean field
Hamiltonian in term of $f$-operators, as defined in equation (\ref{hbfmf}),
is still valid in the present approach.

In agreement with the intuitive picture, the occupation
number derived in the present paper contains two contributions:
one describing itinerant dispersive states with total spectral weight
$Z$ and accounting for the sharp jump at the Fermi energy; and
an incoherent part of localized states with no ${\bf k}$-dependence and
containing $(1-Z)n$ spectral weight. 
The resulting picture is shown in figure 2.

\begin{figure}[t]
\centerline{\psfig{figure=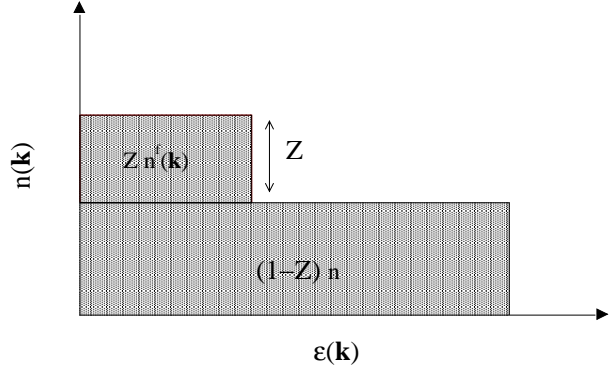,width=8cm}}
\caption{Occupation number of real electrons as calculated
by equation (\ref{nkr}). It is composed by a coherent part with
$Z n^f$ states and an incoherent one with $(1-Z)n$ states.}
\end{figure}

It should be stressed that equation (\ref{ggfl}) 
defines in an unambiguous way
the Green's function of real electrons $G$,
although expressed as function of $G^f$.
It can be now directly employed to calculate any quantity
involving one-particle physics, with no more need to 
deal with auxiliary fields.
In particular, this unique definition of $G$, and
consequently of $n({\bf k})$, yields a consistent
derivation of the total number of electrons and of
the internal energy, overcoming the discrepancy found in the standard
mean field slave-boson solution.

By working directly in $\tilde{c}$-operators, the total
number of electrons $n$ and the internal energy $E$ are expressed
as function of $n_\sigma({\bf k})$ as:
\begin{equation}
n=\sum_{{\bf k},\sigma} \langle \tilde{c}^\dagger_{{\bf k},\sigma}
\tilde{c}_{{\bf k},\sigma}\rangle =\sum_{{\bf k},\sigma} n_\sigma({\bf k}),
\label{nnkc}
\end{equation}
\begin{equation}
E=\sum_{{\bf k},\sigma} 
\left[ t({\bf k})-\mu\right]\langle \tilde{c}^\dagger_{{\bf k},\sigma}
\tilde{c}_{{\bf k},\sigma}\rangle
=\sum_{{\bf k},\sigma} \left[ t({\bf k})-\mu\right] n_\sigma({\bf k}).
\label{fnkc}
\end{equation}
Inserting equation (\ref{nkr}) in (\ref{nnkc}) we then obtain
\begin{equation}
n=\sum_{{\bf k},\sigma} n^f_\sigma({\bf k}).
\label{ntot}
\end{equation}
On the other hand, as previously discussed,
the number of particles can be derived also
by the thermodynamic relation $n = -\partial E / \partial \mu$.
The internal energy, by using the {\em same} expression 
of $n_\sigma({\bf k})$ as in the calculation of $n$, reads:
\begin{equation}
E=\sum_{{\bf k},\sigma} \left[ b^2 t({\bf k})-\mu\right] n^f_\sigma({\bf k}).
\end{equation}

In the same way as in the standard mean field slave-boson derivation,
the physical values of $b^2$ and $\lambda$ that appear
as parameters in $n^f_\sigma({\bf k})$ are found:
\begin{equation}
\lambda = -\sum_{{\bf k},\sigma} t({\bf k})
n^f_\sigma({\bf k}),
\end{equation}
\begin{equation}
b^2=1-\sum_{{\bf k},\sigma} n^f_\sigma({\bf k}).
\end{equation}
Then, deriving $E$ with respect to $\mu$ as required by the
thermodynamic relation, one finally obtains
\begin{equation}
n=\sum_{{\bf k},\sigma} n^f_\sigma({\bf k}),
\end{equation}
just like in equation (\ref{ntot}).

This result shows the fully consistency of the analytic expression
of $G$ derived in the present approach.
The Green's function for real electrons is univocally determined
and its employment to calculate the total number
of electrons and the internal energy
obeys the thermodynamic relation $\partial E / \partial \mu = -n$.
It is also worthy to note that Luttinger's theorem is 
naturally fulfilled by using the Green's function here derived.
This is particularly interesting since the validity of Luttinger's theorem
has not been imposed as a precondition to be satisfied.
This will appear even more relevant if compared with the results
of  mean field theories
performed within the $X$-operator formalism, where not only
the thermodynamic relation $\partial E / \partial \mu = -n$
is not fulfilled, but the validity or the breakdown of Luttinger's
theorem is controversial.
In the next section it will be shown how the inclusion of the incoherent
background, in the same spirit of the present analysis, allows
to overcome all these inconsistencies also in the
mean field solutions based on the Hubbard $X$-operators.

\section{Hubbard operators and mean field theories}

One of the advantages of using $X$-operators is that they
permit to work directly in terms of real electrons, without
introducing any auxiliary field.
In the case of the $t$-model here considered, the $X$-operators
live in the reduced Hilbert space constructed on each site by the states
$|0\rangle$, $|\uparrow\rangle$, $|\downarrow\rangle$.
The Hubbard $X$-operators can then be represented as projection operators:
$X^{\alpha \beta}= |\alpha\rangle\langle\beta |$.
It is easy to check that they follow the algebra:
\begin{equation}
\left[X^{\alpha \beta}_i,X^{\gamma \delta}_j\right]_\pm
=\delta_{ij}\left(\delta_{\beta\gamma}X^{\alpha \delta}_i
\pm \delta_{\alpha\delta}X^{\gamma\beta}_i\right).
\end{equation}
Moreover the more restrictive relation
\begin{equation}
X^{\alpha \beta} X^{\gamma \delta}=\delta_{\beta\gamma}X^{\alpha \delta}
\label{xperx}
\end{equation}
is obeyed.

By using the Hubbard operator formalism, the $t$-model Hamiltonian
can be written as:
\begin{equation}
H=\sum_{i \neq j,\sigma} t_{ij} 
X^{\sigma 0}_i X^{0 \sigma}_j
-\mu\sum_{i,\sigma} X^{\sigma \sigma}_i.
\label{hx}
\end{equation}

In similar way, the Green's function
of real electrons is defined as:
\begin{equation}
G_\sigma(i,j;t)=
-i\langle T_t X^{0 \sigma}_i(t)X^{\sigma 0}_j\rangle.
\label{gxdef}
\end{equation}

Different mean field approximations in terms of Hubbard $X$-operators
have been performed in literature by using several approaches:
diagrammatic studies\cite{izyumov,sandalov}
as well as derivations based on the
equation of motion\cite{plakida}. All these analyses converge to a unique
expression of the mean field Green's function:
\begin{equation}
G_\sigma({\bf k};\omega)=\frac{Z}{\omega-Z t({\bf k})+\mu-\lambda
+\: i{\rm sgn} [Zt({\bf k})-\mu+\lambda]},
\label{gx}
\end{equation}
where, unlike in the slave-boson technique,
\begin{equation}
Z=1-\frac{n}{2},
\end{equation}
and
\begin{equation}
\lambda=-\frac{1}{2}\sum_{{\bf k},\sigma} t({\bf k})
n^f_\sigma({\bf k}).
\end{equation}
Coherently with the spirit of a mean field approximation,
equation (\ref{gx}) describes a system of non interacting electrons
with total spectral weight $(1-n/2)$ and with a dispersive 
band renormalized by the same factor.

The occupation number is directly evaluated by (\ref{gx}).
In fact, equation (\ref{xperx}) implies as a particular case the relation
\begin{equation}
X^{\sigma \sigma}_i = X^{\sigma 0}_i X^{0 \sigma}_i,
\end{equation}
hence
\begin{equation}
n_\sigma({\bf k})=\left(1-\frac{n}{2}\right)
\Theta\left[\mu-\left(1-n/2\right) t({\bf k})-\lambda\right],
\label{nxk}
\end{equation}
where $\Theta$ is the Heaviside function.
The corresponding total number of electrons,
\begin{equation}
n=\left(1-\frac{n}{2}\right) \sum_{{\bf k},\sigma}n_\sigma({\bf k}),
\label{nx}
\end{equation}
shows one the peculiarities of the mean field solutions based
on $X$-operators, namely the breakdown of Lut\-tin\-ger's theorem.
In fact, according to equation (\ref{nx}), the half-filled case $n=1$
corresponds to the complete filling of the electronic band
and to a vanishing Fermi surface.
On the contrary, the half-band filling case corresponds to
$n=2/3$ leading to unphysical results. For instance, all the
properties and instabilities of the system  due to possible
nesting of the Fermi surface, as antiferromagnetic or
charge-den\-si\-ty-wa\-ve ordering, lie around this particular value
$n=2/3$, a clear artifact of the approximation.

On the other hand, the same breakdown of Luttinger's theorem
appears ques\-tion\-able. One could as well use the above widely
discussed thermodynamic relation $\partial E / \partial \mu = -n$
to determine the total number of electrons.
However, it is easy to check that the expression (\ref{nxk}),
once plugged in the internal energy
\begin{equation}
E=\sum_{i \neq j,\sigma} \left[t_{ij} -\mu\delta_{ij}\right]
\langle X^{\sigma 0}_i X^{0 \sigma}_j\rangle
=
\sum_{{\bf k},\sigma}
\left[ t({\bf k})-\mu\right]n_\sigma({\bf k}),
\label{fx}
\end{equation}
does not reproduce the result of equation (\ref{nx}).

We found ourselves in the same controversial situation as for
the slave-boson solution. As before, one should postulate that the
electronic propagator in equation (\ref{gx}) (and related functions
like spectral function or occupation number $n_\sigma({\bf k})$)
has to be used only for non local quantities as the kinetic energy,
but it does not give any information about the total number of particles.
This inconsistency was in that case solved by the introduction
of the incoherent background.
An identical result will be also recovered in the Hubbard operator approach.

Let us rewrite the ``correct'' occupation number by adding
an incoherent con\-tri\-bu\-tion $B$:
\begin{equation}
n_\sigma({\bf k})=\left(1-\frac{n}{2}\right)
\Theta\left[\mu-\left(1-n/2\right) t({\bf k})-\lambda\right]+B.
\label{nxk2}
\end{equation}
The total number of electrons is thus given by:
\begin{equation}
n=\left(1-\frac{n}{2}\right) \sum_{{\bf k},\sigma}
\Theta\left[\mu-\left(1-n/2\right) t({\bf k})-\lambda\right]
+2B.
\label{nx2}
\end{equation}
By substituting equation (\ref{nxk2}) in (\ref{fx}), the internal energy
becomes:
\begin{equation}
E=
\left(1-\frac{n}{2}\right)\sum_{{\bf k},\sigma}
t({\bf k})
\Theta\left[\mu-\left(1-n/2\right) t({\bf k})-\lambda\right]
-\mu n.
\label{fx2}
\end{equation}
The thermodynamic relation can now be employed to determine
the spectral weight of the incoherent background $B$ by requiring
the total number of electrons to be equal to equation (\ref{nx2}).
The derivative of the internal energy in equation (\ref{fx2}) with respect to
$\mu$ gives
\begin{equation}
\frac{\partial E}{\partial \mu}=-n
=
\frac{\partial n}{\partial \mu}\lambda
-n -\mu\frac{\partial n}{\partial \mu}
+\left(\mu-\lambda\right)\frac{\partial}{\partial \mu}
\sum_{{\bf k},\sigma}\Theta\left[\mu-\left(1-n/2\right) 
t({\bf k})-\lambda\right],
\label{derf}
\end{equation}
where the equality
\begin{equation}
\left(1-\frac{n}{2}\right)t({\bf k})
\frac{\partial}{\partial \mu}
\Theta\left[\mu-\left(1-n/2\right)t({\bf k})-\lambda\right]
=\left(\mu-\lambda\right)
\frac{\partial}{\partial \mu}
\Theta\left[\mu-\left(1-n/2\right)t({\bf k})-\lambda\right]
\end{equation}

has been used.
Equation (\ref{derf}) can be further simplified to obtain
\begin{equation}
\frac{\partial n}{\partial \mu}=
\frac{\partial}{\partial \mu}
\sum_{{\bf k},\sigma}
\Theta\left[\mu-\left(1-n/2\right)t({\bf k})-\lambda\right],
\end{equation}
whose solution is trivially:
\begin{equation}
n=\sum_{{\bf k},\sigma}
\Theta\left[\mu-\left(1-n/2\right)t({\bf k})-\lambda\right].
\label{nx3}
\end{equation}
The constant $B$ is thus easily obtained by equating
(\ref{nx2}) with (\ref{nx3}):
\begin{equation}
B=\frac{n^2}{4}.
\end{equation}
Note that the validity of Luttinger's theorem is now
unambiguously defined and, just as in the slave-boson formalism,
arises in a natural way from the correct $n_\sigma({\bf k})$
which includes the proper incoherent background.

\section{Summary and discussion}

A slightly modified version of the slave-boson mean field
approximation has been proposed in this paper to account for the localized
incoherent background missing in the standard mean field theories.
In contrast to them,
the resulting Green's function here derived can be safely used in the
calculation of any, local or non local, quantity as the kinetic energy,
the total number of electrons or the internal energy.
It has also been shown that Luttinger's theorem is naturally fulfilled
when such a Green's function is employed.

This approach permits a particular compact and suitable way,
although ap\-prox\-i\-mate, to deal with the redistribution of the 
one-particle spectral weight due to strong correlation effects.
Broken symmetry states, at a mean field level,
can be also taken into account in similar way.
A straightforward generalization is the superconducting case.
It is well known that no purely local $s$-wave Cooper pair
can be established in a strongly correlated system, once
an attractive interaction is taken into account, because of the
forbidding of double occupancy. However, a common shortcoming
of the standard mean field theories is to allow for a finite
local $s$-wave order parameter as a consequence of
the relaxing of the local constraint\cite{plakida}.
This spurious result disappears
when the modified mean field approximation here proposed
is used. Just in the same way as in the normal state, the
anomalous propagator $F(ij;t)=-i\langle 
\tilde{c}^\dagger_{i\uparrow}(t)\tilde{c}^\dagger_{j\downarrow}\rangle$
can be splitted in a local and non local part. Following
the previous procedure, it easy to obtain the corresponding
BCS expression for the anomalous Green's function $F$:
\begin{equation}
F({\bf k};\omega)= - Z 
\frac{\Delta({\bf k})}{\omega^2-\xi^2({\bf k})-\Delta^2({\bf k})}
+Z \sum_{{\bf k}} 
\frac{\Delta({\bf k})}{\omega^2-\xi^2({\bf k})-\Delta^2({\bf k})}.
\label{ffl}
\end{equation}
Note that both the local part has the same spectral weight $Z=b^2$ as the
non local one because the relation (\ref{bfi}) can not be employed in this case.
and the non local part have the same spectral weight.
From equation (\ref{ffl}) it is thus clear that the condensate relative to
local Cooper pairs, given by
the order parameter $\sum_{{\bf k}} F({\bf k};\omega)$, is
identically zero, as physically expected.

\nonumsection{Acknowledgements}
\noindent
The author would like to thank R. Zeyher and M. Grilli
for the useful discussions.

\end{document}